\documentclass[aps,prl,showpacs,twocolumn,oneside,english,superscriptaddress,floatfix]{revtex4}

\usepackage{amssymb}
\usepackage{natbib}
\usepackage{graphicx}
\usepackage{amsmath}
\usepackage{wasysym}
\usepackage[bookmarks = false]{hyperref}
\usepackage{color}

\begin{document}

\title{An efficient quantum light-matter interface with sub-second lifetime}

\author{Sheng-Jun Yang}
\affiliation{Hefei National Laboratory for Physical Sciences at Microscale and Department
of Modern Physics, University of Science and Technology of China, Hefei,
Anhui 230026, China}
\affiliation{CAS Center for Excellence and Synergetic Innovation Center in Quantum
Information and Quantum Physics, University of Science and Technology of
China, Hefei, Anhui 230026, China}
\author{Xu-Jie Wang}
\affiliation{Hefei National Laboratory for Physical Sciences at Microscale and Department
of Modern Physics, University of Science and Technology of China, Hefei,
Anhui 230026, China}
\affiliation{CAS Center for Excellence and Synergetic Innovation Center in Quantum
Information and Quantum Physics, University of Science and Technology of
China, Hefei, Anhui 230026, China}
\author{Xiao-Hui Bao}
\affiliation{Hefei National Laboratory for Physical Sciences at Microscale and Department
of Modern Physics, University of Science and Technology of China, Hefei,
Anhui 230026, China}
\affiliation{CAS Center for Excellence and Synergetic Innovation Center in Quantum
Information and Quantum Physics, University of Science and Technology of
China, Hefei, Anhui 230026, China}
\author{Jian-Wei Pan}
\affiliation{Hefei National Laboratory for Physical Sciences at Microscale and Department
of Modern Physics, University of Science and Technology of China, Hefei,
Anhui 230026, China}
\affiliation{CAS Center for Excellence and Synergetic Innovation Center in Quantum
Information and Quantum Physics, University of Science and Technology of
China, Hefei, Anhui 230026, China}

\begin{abstract}
Quantum repeater holds the promise for scalable long-distance quantum communication. Towards a first quantum repeater based on memory-photon entanglement, significant progresses have made in improving performances of the building blocks. Further development is hindered by the difficulty of integrating key capabilities such as long storage time and high memory efficiency into a single system. Here we report an efficient light-matter interface with sub-second lifetime by confining laser-cooled atoms with 3D optical lattice and enhancing the atom-photon coupling with a ring cavity. An initial retrieval efficiency of 76(5)\% together with an 1/e lifetime of 0.22(1) s have been achieved simultaneously, which already support sub-Hz entanglement distribution up to 1000 km through quantum repeater. Together with an efficient telecom interface and moderate multiplexing, our result may enable a first quantum repeater system that beats direct transmission in the near future.
\end{abstract}

\maketitle

\vspace{0.5cm}

Quantum communication~\cite{Gisin2007,Kimble2008} relies on photon transimission over long distance. Direct transmission is limited to moderate distances (less than 500~km~\cite{Gisin2015,Korzh2014}) due to exponential decay of photons. Quantum repeater~\cite{Briegel1998} is an ultimate solution to go significantly beyond this limitation. There are many quantum repeater schemes proposed so far~\cite{Sangouard2011,Munro2010,Fowler2010,Munro2012,Muralidharan2014,Azuma2015}. Considering the experimental capabilities, the memory-photon entanglement based schemes~\cite{Sangouard2011} are much more feasible than the error correction based schemes~\cite{Sangouard2011,Munro2010,Fowler2010,Munro2012,Muralidharan2014,Azuma2015}. Towards a first quantum repeater based on memory-photon entanglement, significant progresses have made in improving performances of the building blocks~\cite{Sangouard2011}. Further development is hindered by the difficulty of integrating key capabilities such as long storage time~\cite{Radnaev2010} and high memory efficiency~\cite{Simon2007} into a single system. So far, storage lifetime has been improved to the sub-second regime for single excitations in an atomic ensemble~\cite{Radnaev2010,Dudin2010prl} and to the sub-minute regime for classical light storage~\cite{Dudin2013}, nevertheless, storage efficiency in these experiments is typically very low ($\sim$16\%). If we define a threshold of 50\% for the memory efficiency, the longest storage time is limited to 1.2\,ms~\cite{Bao2012}, which is far away from the second regime requirement~\cite{Sangouard2011} of a quantum repeater. Memory efficiency is essentially important as it intervenes in every entanglement swapping operation between adjacent quantum repeater nodes. According to theoretical estimations~\cite{Sangouard2011}, 1\% increase of retrieval efficiency can improve long-distance entanglement distribution rate by 10\%$\sim$14\%.

In this paper, we report an efficient quantum memory with sub-second regime lifetime by making use of a 3D optical lattice confined atomic ensemble inside a ring cavity. The quantum memory is nonclassically correlated with a single photon, thus forms a light-matter interface for quantum repeaters~\cite{Duan2001,Sangouard2011}. Optical lattice limits atomic motion in all direction thus suppresses various motion-induced decoherence, and ring cavity enhances atom-photon coupling thus improves the retrieval efficiency. We further use the magic trap technique~\cite{Lundblad2010,Dudin2010pra} to compensate the lattice induced differential light shift. To be compatible with sub-second regime storage, the ring cavity is stabilized with a large-detuned reference beam. By taking all these measures, we finally realize a light-matter interface with an initial efficiency of 76(5)\% and a $1/\rm{e}$ lifetime of 0.22(1)\,s. This is a significant step forward in realizing a high-performance light-matter interface filling the requirement of quantum repeaters.

\begin{figure}[h]
\includegraphics[width=\columnwidth]{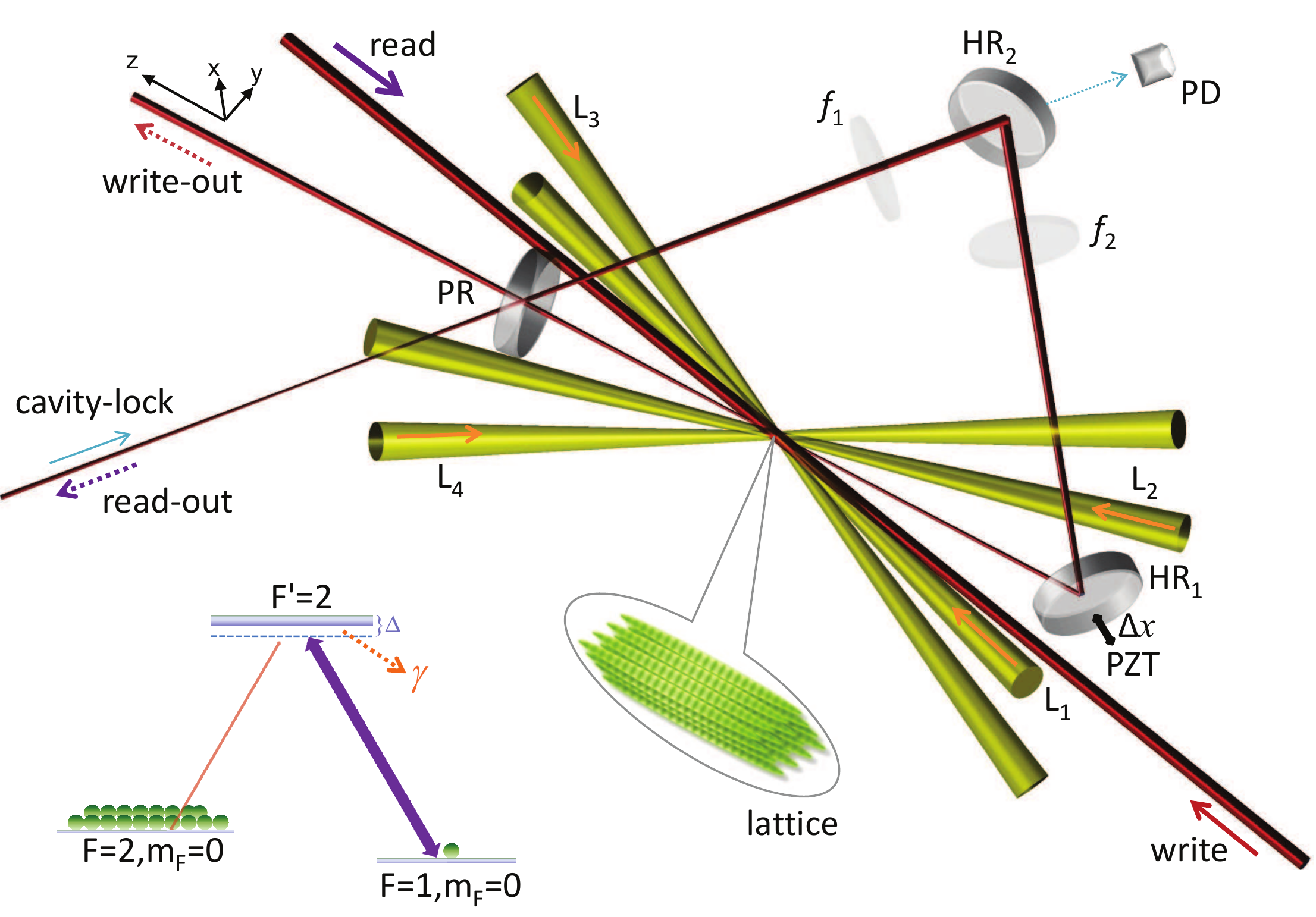}
\caption{Experimental setup and atomic levels. The $^{87}\rm{Rb}$ atoms are trapped with a 3D optical lattice by interfering four circularly polarized 1064\,nm laser beams ($\mathrm{L}_i$, $i\!=\!\{1,2,3,4\}$). The atoms are initially prepared in the ground state $\left|F\!=\!2,m_F\!=\!0\right>$. Another ground state of $\left|F\!=\!1,m_F\!=\!0\right>$ is employed for storage. The ring cavity, in the horizontal plane (the yz-plane), consists of one partially reflecting mirror PR of 92.0(3)\% reflection rate, two highly reflecting mirrors $\rm{HR}_1$ and $\rm{HR}_2$, and two plano-convex lenses $f_1$ and $f_2$ with a same focus length of 250\,mm. The orthogonal linear polarized write and read pulses of the DLCZ memory are counter-propagating through the atoms, having a separation angle of $\theta=3^{\circ}$ with respect to the cavity mode. The write-out and read-out photons are cavity-enhanced and escape from the two coupling ports of the mirror PR. The cavity-locking beam is combined into the read-out channel with a polarized beam splitter. Leaks of the locking pulses from the mirror $\rm{HR}_{2}$ are detected by a fast photodiode (PD), and fed forward to stabilize the cavity length by displacing the mirror $\rm{HR}_{1}$ with a piezoelectric transducer (PZT).}
\label{fig:Setup}
\end{figure}

The experimental setup is shown in Fig.~\ref{fig:Setup}. An ensemble of $^{87}\!\rm{Rb}$ atoms is first prepared through magneto-optical trapping~(MOT) and cooled down to about $12\,\rm{\mu K}$ via optical molasses. The atoms are finally loaded into a 3D optical lattice via interfering four 1064-nm laser beams at the setup centre, and optically pumped to the ground state $\left|F\!=\!2, m_{F}\!=\!0\right\rangle$. The optical lattice is formed by four circularly polarized laser beams $\rm{L}_{\emph{i}}$ ($i\!=\!\{1,2,3,4\}$), with the same beam waist diameter of $500\,\rm{\mu m}$ overlapping through the atoms, and $12^\circ$ axisymmetric angled to the bias magnetic field $B$. The 1064\,nm laser output is $50\%\!:\!50\%$ split into two beams $\rm{L}_{1}$ and $\rm{L}_{2}$, then reflected backward and used as the beams $\rm{L}_{3}$ and $\rm{L}_{4}$ respectively. As the vacuum glass cell is not anti-reflecting coated for 1064\,nm, power of $\rm{L}_3$ ($\rm{L}_4$) is $19\%$ lower than $\rm{L}_1$ ($\rm{L}_2$). Lattice periods are 5.9\,$\rm{\mu m}$, 2.8\,$\rm{\mu m}$ and 0.54\,$\rm{\mu m}$ for the x, y and z directions respectively. The lattice-trapped ensemble is 0.2\,mm wide in the xy plane and 0.8\,mm long in the z direction. The optical depth is about 1.6 in the z direction. A temporal gap of 90\,ms is left in the lattice phase to wait for all the unbounded atoms flying away. The write pulse is vertically polarized, of pulse width 120\,ns and beam power $3.8\,\rm{\mu W}$; while the read pulse is horizontally polarized, of pulse width 240\,ns and beam power $330\,\rm{\mu W}$. The write-out photons are collected by a single-mode fibre, and measured by single-photon detector $D_1$ after a Fabry-Perot cavity filter; the read-out photons are collected and sent through a vapour cell atomic filter, and then measured by single-photon detectors $D_2$ and $D_3$ after $50\%\!:\!50\%$ splitting. In order to suppress the beam leakages and their influence on the stored single excitations in the ensemble, most of the laser beams are controlled by double-passed acousto-optic modulators and the laser beams for the MOT are blocked with a mechanical chopper during memory operations.

\begin{figure}[h]
\includegraphics[width=\columnwidth]{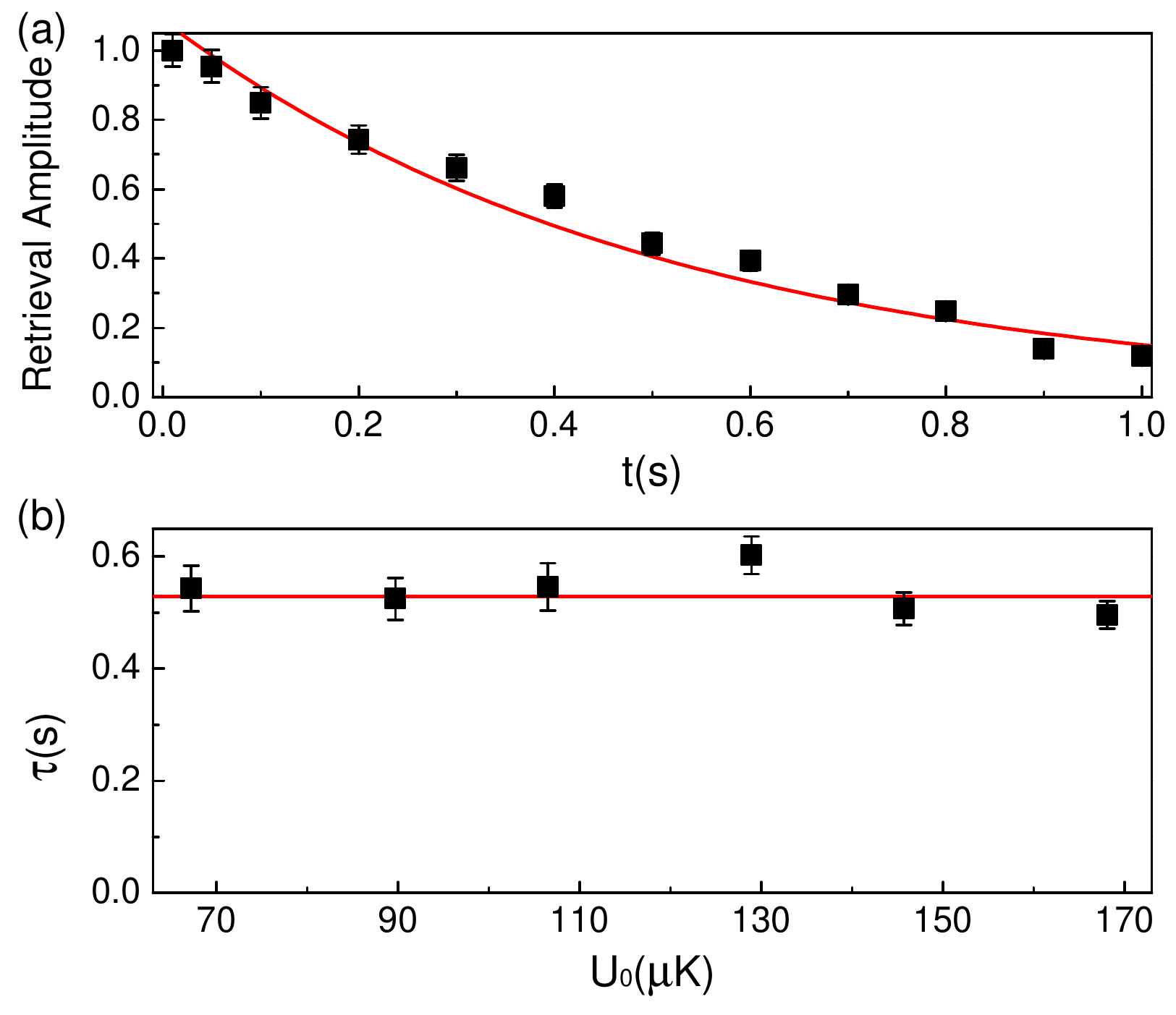}
\protect\caption{Lifetime measurement of the optical lattice-trapped EIT memory without a cavity. (a) Retrieval amplitude normalized to the point of $t=10\,{\rm{ms}}$. Fitting with an exponential function (red line), we get an $1/e$ lifetime of $\tau=0.51(3)\,\rm{s}$. Lattice potential is set to $U_0=146\,{\rm \mu K}$ for this measurement. (b) Lifetime measured for different trap potentials. The average value is $\tau=0.53\,\rm{s}$ (red line).}
\label{fig:EIT}
\end{figure}

Confinement of atom motion with a 3D optical lattice suppresses all motion-induced decoherence\cite{Zhao2008}, but gives rise to a new decoherence of differential light shift. By adding an appropriate bias magnetic field $B$, differential light shift from the vector polarizability compensates that from the scalar part~\cite{Lundblad2010}. Without the ring cavity, we first calibrate the magic condition via the electromagnetically induced transparency (EIT) memory~\cite{Fleischhauer2005}. The single-photon-level probe pulse to be stored is resonant with the D1 line $\left|F\!=\!2\right>\!\to\!\left|F'\!=\!2\right>$ and transmits in the write-out direction in Fig.~\ref{fig:Setup}. The control pulse is resonant with the transition $\left|F\!=\!1\right>\!\to\!\left|F'\!=\!2\right>$ and transmits in the write direction. The probe and control pulses are controlled to enable a stopped-light configuration~\cite{Fleischhauer2005}. By optimizing the retrieval amplitude at $t\!=\!0.5$\,s, we find the best magic compensating field to be $4.56\,\rm{Gauss}$. With a lattice potential of  $U_{0}\!=\!146\,\rm{\mu K}$, we measure the retrieval signal decay as a function of storage time in Fig.~\ref{fig:EIT}\,(a). The $1/e$ time constant is fitted to be 0.51(3)\,s. Changing the lattice laser power $U_{0}$, the storage lifetime are more or less the same, with an average value of 0.53\,s (Fig.~\ref{fig:EIT}\,(b)). This indicates that the differential light shift has been successfully eliminated.

Afterwards, we set up a ring cavity around the lattice-trapped atoms to enhance the atom-photon coupling strength and carry out the DLCZ~\cite{Duan2001} memory with single photons. The orthogonal linear-polarized write and read pulses are counter-propagating through the atoms (Fig.~\ref{fig:Setup}). The write and read beams are red detuned by 40\,MHz to the D1 transitions. The write-out and read-out photons are configured to be resonant with the ring cavity. The cavity has a finesse of $\mathcal{F}\!=\!52$, and a linewidth of 9.2~MHz. In our previous experiments~\cite{Bao2012}, the cavity was intermittently stabilized to avoid influence from the locking beam on to the memory. In order to obtain a high retrieval efficiency for long storage durations, we have developed the cavity stabilization technique with large detuning of atomic ransitions. The cavity locking beam is from a reference laser of $\lambda\!=\!800\,\rm{nm}$ with 5\,nm detuned to the D1 line transitions, and is stabilized with an ultra-stable cavity due to the unavailability of atomic lines. The cavity locking beam has a power of $1\,\rm{\mu W}$, which introduces a differential light shift of 0.3 Hz in side the cavity. In our experiment, we turn on the locking beam for 10\,ms to pull back the cavity resonance before applying the read pulse.

\begin{figure}[h]
\includegraphics[width=\columnwidth]{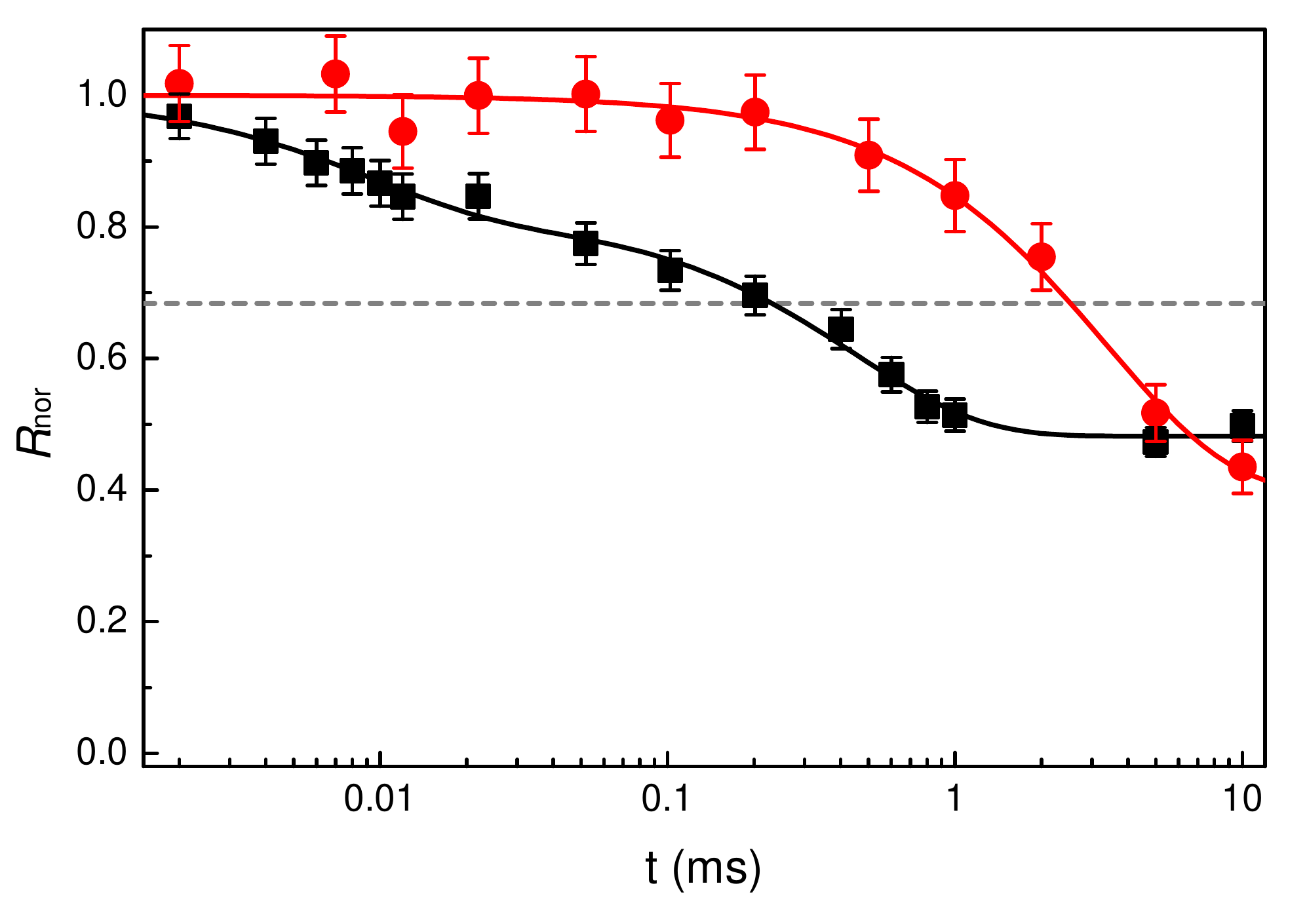}
\caption{Millisecond regime decay of retrieval efficiency with DLCZ storage. Retrieval efficiency $R$ is normalized to the initial value at $t=0$. Data points in black square refer to the angled case $\theta=3^{\circ}$, while points in red dot refer to the collinear case $\theta=0^{\circ}$. For a reference of $R_{\rm{nor}}=0.5e^{-1}+0.5$ (gray dashed line), the decay time is 0.23\,ms for the angled case, while 2.52\,ms for the collinear case. The lattice potential $U_0$ has a fixed value of $70\,\rm{\mu K}$.}
\label{fig:Dephasing}
\end{figure}

By applying a write pulse, a write-out photon is detected with an overall probability of $p_{\rm{w}}\!\sim\!10^{-3}$ and heralds the creation of a collective excitation in the atomic ensemble. Since $p_{\rm{w}}$ is low, we repeat the write trials till a write-out signal photon is detected. The retrieval process with an adjustable delay only starts after a successful write-out event. Retrieval efficiency is measured as $R=p_{\rm{r|w}}-p_{\rm{r}}$, where $p_{\rm{r}|\rm{w}}$ ($p_{\rm{r}}$) refers to the conditional (unconditional) detection probability of a read-out photon.
Unexpectedly, we observe significant efficiency decay within the initial sub-millisecond regime as shown in Fig.~\ref{fig:Dephasing} in black. We suspect that this is due to imperfect lattice potentials which result in some free atoms not confined within single lattice sites. Thus we make the same measurement for the collinear case~\cite{Zhao2008} ($\theta=0$), with the result shown in Fig.~\ref{fig:Dephasing} in red. The retrieval efficiency drops by about 50\% in both measurements but with different time scales. The angled case drops faster since the phase grating gets distorted for transversal movement of a half spinwave wavelength ($\lambda_{\rm{sw}}/2=7.6\,\rm{\mu m}$). While in the collinear case, since the spinwave wavelength is largely increased to 4\,cm, transversal motion influences the retrieval efficiency by distorting the transversal mode of the read-out photon, which is sensitive to the movement scale of the cavity mode waist ($w_{\rm{cav}}=60\,\rm{\mu m}$). Considering an atom temperature of 12\,$\rm{\mu K}$, the time required for movement over $\lambda_{\rm{sw}}/2$ or $w_{\rm{cav}}$ coincide with the decay time observed in Fig.~\ref{fig:Dephasing}. In order to minimize these free atoms, we dynamically increase $U_0$ from $70\,\rm{\mu K}$ to $180\,\rm{\mu K}$ after all the unloaded atoms flying out the lattice region. By doing this, the dropping rate of retrieval efficiency gets reduced remarkably to $1\!-\!R(10\,\mathrm{ms})/R(0\,\mathrm{\mu s})\!\approx\!20\%$, which is a significant improvement compared to previous experiments with 1D lattice~\cite{Zhao2009,Radnaev2010,Dudin2010prl} and hollow beam dipole trap~\cite{Yang2011}.

\begin{figure}[h]
\includegraphics[width=\columnwidth]{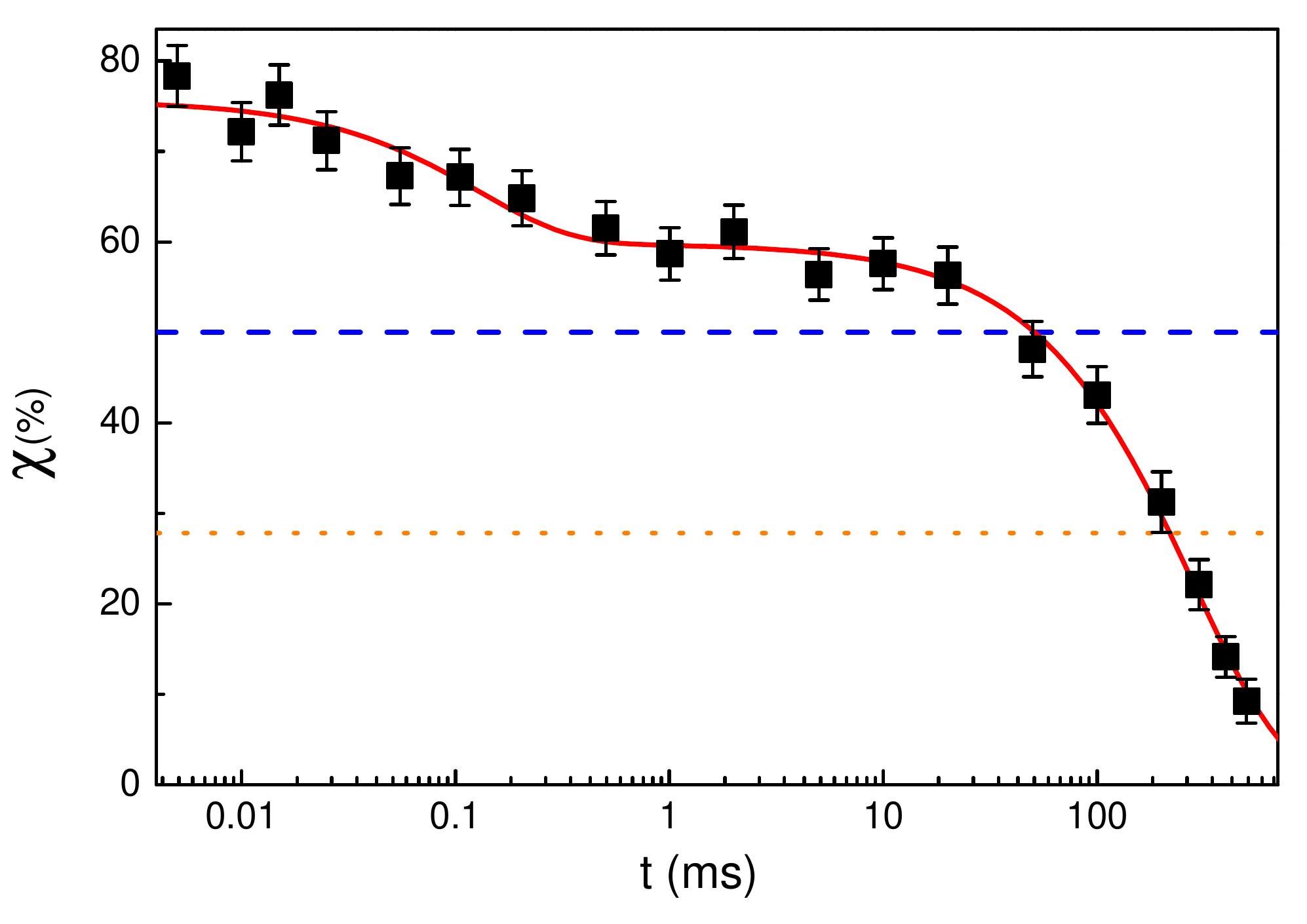}
\caption{Intrinsic retrieval efficiency $\chi$ versus storage time for DLCZ storage. Data points are fitted with a double-exponential decay function (red line). The initial intrinsic retrieval efficiency is $\chi(0)=76(5)\%$.
For a threshold of $\chi=50\%$ (blue dashed line), our memory persists until 51\,ms. While for a threshold of $\chi=e^{-1}\chi(0)$ (orange dotted line), our memory persists until 0.22\,s. }
\label{fig:DLCZ}
\end{figure}

In Fig.~\ref{fig:DLCZ} we show the final result of retrieval efficiency measurement in the range of 0$\sim$500\,ms. The intrinsic retrieval efficiency $\chi$ is obtained through $\chi=R/(\eta _{\mathrm{cav}}\,\eta _{\mathrm{t}}\,\eta _{\mathrm{spd}})$, where $\eta _{\mathrm{cav}}=68(3)\%$ refers to the photon emanation rate from the ring cavity, $\eta _{\mathrm{t}}=55(2)\%$ refers to the propagation efficiency from the cavity to the detectors, and $\eta _{\mathrm{spd}}=63(2)\%$ refers to the detection efficiency for the single-photon detectors. We fit the data with a double-exponential decay function $\chi\left(t\right)\!=\!\chi_1\exp\left(-t/\tau_1\right)\!+\!\chi_2\exp\left(-t/\tau_2\right)$, where $\tau_1$ and $\tau_2$ are the two characteristic decay parameters. The fitted parameters are $\tau_1=$0.13(4)\,ms, $\tau_2$=285(12)\,ms, $\chi_1$=$15.8\pm1.8$\% and $\chi_2$=$59.8\pm4.0$\% respectively.
The initial intrinsic retrieval efficiency is $\chi(0)=76(5)\%$, which is comparable with the best previous results on efficiency in the single-quanta regime~\cite{Simon2007, Bao2012}. The $1/e$ decay time for the retrieval efficiency is $0.22(1)\,\rm{s}$, which is also comparable with the best previous results on lifetime in the single-quanta regime~\cite{Radnaev2010,Dudin2010prl}. For an threshold of $\chi\geq50$\%, our memory can persist until $51\,\rm{ms}$, while the best previous result is merely 1.2\,ms~\cite{Bao2012}. We would like to stress that efficient storage far beyond the millisecond regime is essentially important for quantum repeaters~\cite{Sangouard2011}. To further identify that our experiment genuinely operates in the single-quanta regime, we measure the anticorrelation parameter~\cite{U'Ren2005} $\alpha$ for different storage times, with the result shown in Table~\ref{tab:alpha}. The write-out photon is detected with $D_{1}$, while the read-out photon is 50:50 split and detected with two separate detectors $D_{2}$ and $D_{3}$. The $\alpha$ parameter is defined as
\begin{equation}
\alpha\!=\!p_{23|1}/(p_{2|1} \cdot p_{3|1}),
\end{equation}
where $p_{j|i}$ ($p_{jk|i}$) is the conditional single (two-fold coincidence) detection probability. The measured $\alpha$ parameter at $t=0\,\rm{s}$ and $t=0.5\,\rm{s}$ are $0.11(5)$, and $0.30(21)$ respectively, both of which are well below the classical threshold of $\alpha\geq1$, implying the non-classical behaviour of our memory.

\begin{table}[h]
\caption{Measurement of the $\alpha$ parameter.}
\begin{tabular*}{\columnwidth}{@{\extracolsep\fill}cccccc}\toprule
$t$(ms) & $\alpha$ & $n_{123}$ & $R$(\%) & $\chi$(\%) & $T_{\rm{m}}$(hour)\\\hline
0 &0.11(5) & 6$<$55 & 17.1(5) & 73(5) & 1.4\\
10 &0.16(7) & 6$<$37 & 12.4(4) & 53(4) & 5.3\\
100 &0.28(11) & 7$<$25 & 9.7(3) & 41(3) & 6.1\\
300 &0.09(9) & 1$<$11 & 5.4(2) & 23(2) & 8.6\\
500 &0.30(21) & 2$<$7 & 2.6(1) & 10(1) & 24.7\\
\toprule\end{tabular*}
\small\begin{flushleft} Notes: $t$, storage time; $n_{123}$, number of triple coincidence events of the detectors $D_1$, $D_2$ and $D_3$; $R$, measured retrieval efficiency; $\chi$, intrinsic retrieval efficiency; $T_{\rm{m}}$, measurement duration. For the column of $n_{123}$, the right part of each inequality refers to the expected triple coincidence when $\alpha=1$. \end{flushleft}
\label{tab:alpha}
\end{table}

Realization of a light-matter interface with high efficiency and long lifetime has significant applications in long-distance quantum communication. If we consider a DLCZ quantum repeater with multiplexing~\cite{SimonC2007,Collins2007,Sangouard2011}, our current result on efficiency and lifetime already supports sub-Hz entanglement distribution up to 1000 km~\cite{SuppleNote}, assuming channel loss of 0.16 db/km~\cite{Korzh2014}, perfect telecom interface, single-photon detectors with 100\% efficiency and moderate multiplexing. Our work thus demonstrates for the first time that efficiency and lifetime meet the requirement of quantum repeater simultaneously. Higher intrinsic retrieval efficiency will be possible by using a cavity with higher finesse. Longer storage time will be possible by using dynamical decoupling~\cite{Dudin2013}. Besides, as the optical lattice tightly confines atoms in all directions, our system has a feasible spatial multiplexing capacity~\cite{Tordrup2008, Surmacz2008} with long lifetime. In order to realize a practical quantum repeater, coupling losses have to be minimized and telecom interface with high efficiency should be integrated. Our result will also be very useful for creating large-scale cluster states for one-way quantum computing~\cite{Barrett2010}.

\begin{acknowledgments}
This work was supported by the National Natural Science Foundation of China, National Fundamental Research Program of China (under Grant No. 2011CB921300), and the Chinese Academy of Sciences. X.-H.B. acknowledge support from the Youth Qianren Program.
\end{acknowledgments}

\appendix

\section{Supplementary Information}

\begin{figure}[htb]
\includegraphics[width=\columnwidth]{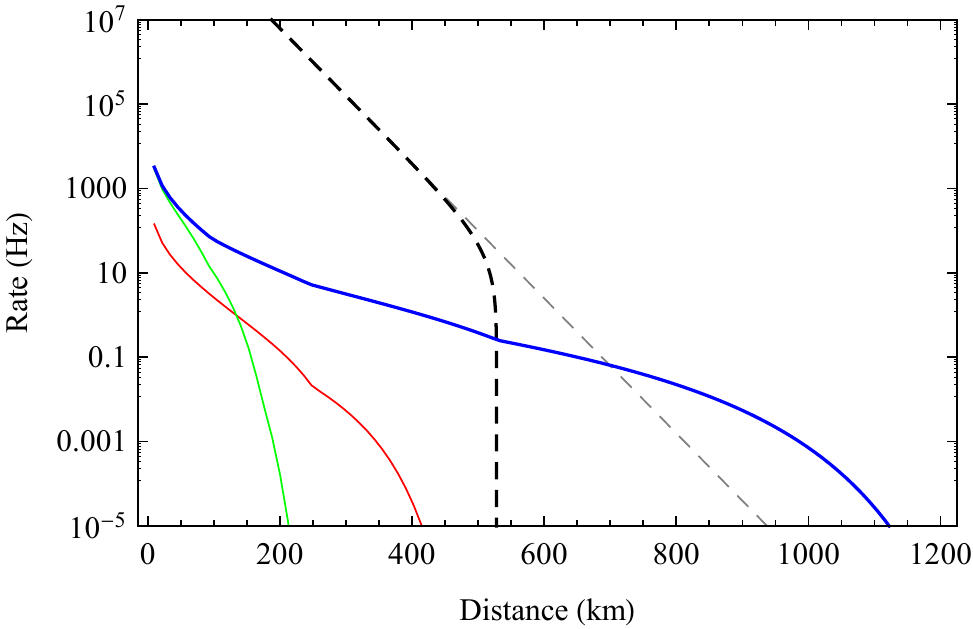}
\caption{Entanglement distribution rate for quantum repeater versus single-photon distribution rate through direct transmission. Dashed lines correspond direct transmission (Gray: no dark count, Black: dark count probability of $10^{-9}$ considered). Solid lines correspond to quantum repeater with different memory parameters (lifetime + efficiency). Red: 0.2~s + 16\% (Radnaev \textit{el al.} in 2010~\cite{Radnaev2010}), Green: 3.2~ms + 73\% (Bao \textit{et al.} in 2012~\cite{Bao2012}), Blue: 0.22~s + 76\% (result of this paper).}
\label{fig:repeater}
\end{figure}

Below we present calculations of entanglement distribution rate through quantum repeater and make comparisons with direct transmission~\cite{Sangouard2011}. For direction transmission, we consider ideal single photons with a repetition rate of 10~GHz and using low-loss fibers with 0.16~dB/km and detectors of 100\% efficiency. The calculation result is shown in Fig.~\ref{fig:repeater} as dashed lines. If dark count is not considered, the distribution rate simply drops exponentially. By choosing a threshold distribution rate of $R=0.01~\rm{Hz}$, the maximal distance is 750~km. When a reasonable dark count probability of $10^{-9}$ is considered, the distribution rate gets cutoff for a distance of 530~km. We note that the cutoff distance for direct transmission may vary slightly by assuming different parameters~\cite{Gisin2015}.

For quantum repeaters, we consider the DLCZ protocol~\cite{Duan2001} with multiplexing~\cite{Sangouard2011,SimonC2007,Collins2007}. We optimize the nesting level $n$ and set the limit of $n\leq3$. We assume using the same fiber and detectors as direct transmission. We also assume perfect telecom interface. Pair emission probability $p$ and multiplexing number $N_m$ are selected such that $N_m p=1$ and the fidelity due to multi-photon errors is 95\%. We note that low dark count probability of $10^{-9}$ has negligible influence in quantum repeater with small nesting levels~\cite{Brask2008} since entanglement connection and detection are conducted heraldedly. For the quantum memory parameters (lifetime + efficiency), we consider three representative results in the single quantum regime with cold atomic ensembles, which includes the result of 0.2~s + 16\% by Radnaev \textit{el al.} in 2010~\cite{Radnaev2010}, the result of 3.2~ms + 73\% by Bao \textit{et al.} in 2012~\cite{Bao2012} and the current result of 0.22~s + 76\% in this paper. We calculate the generation rate of a pair of two-photon entanglement over distance $L$ for the three groups of memory parameters. In our calculation, a memory decay model of $e^{-t/\tau}$ is assumed for simplicity. The results are shown as solid lines in Fig.~\ref{fig:repeater}. The multiplexing number $N_m$ varies from 40 up to 1000 for different distances and different memory parameters. It is clear that the previous memories merely allow entanglement distribution with similar or even worse scaling than direct transmission. For a threshold $R=0.01~\rm{Hz}$, the maximal distance is 280~km, which is not comparable with direction transmission. In contrast, our new result shows a much better scaling than direction transmission and enables entanglement distribution much longer than the cutoff distance of direct transmission. For a threshold $R=0.01~\rm{Hz}$, our new result supports a quantum repeater over 860~km. If the memories at the two terminals are retrieved and detected immediately after storage (valid for quantum key distribution), the effective entanglement distribution rate goes much higher (0.04~Hz @1000~km).

We conclude that our new result on memory efficiency and lifetime for the first time enables to build a quantum repeater system that beats direct transmission.

\bibliography{myref}

\end{document}